\documentclass[journal,twocolumn,final]{IEEEtran}

\ifCLASSINFOpdf
\else
\fi
\usepackage{soul} 

\usepackage[cmex10]{amsmath}
\usepackage{amssymb,amsfonts,amsthm,graphicx,cite}
\DeclareGraphicsExtensions{.pdf,.jpeg,.png}
\usepackage{booktabs}
\usepackage{algorithmic}

\usepackage{array}

\usepackage{balance}

\begin{document}

\title{Diagonal Unloading Beamforming \protect\linebreak for Source Localization}

\author{Daniele~Salvati,  
				~Carlo~Drioli,
        and~Gian~Luca~Foresti, 
\thanks{
D. Salvati, C. Drioli, and G.L. Foresti are with the Department
of Mathematics and Computer Science, University of Udine, Udine 33100, Italy, e-mail: daniele.salvati@uniud.it, carlo.drioli@uniud.it, gianluca.foresti@uniud.it.}
}

{

\maketitle

\begin{abstract}
In sensor array beamforming methods, a class of algorithms commonly used to estimate the position of a radiating source, the diagonal loading of the beamformer covariance matrix is generally used to improve computational accuracy and localization robustness.    
This paper proposes a diagonal unloading (DU) method which extends the conventional response power beamforming method by imposing an additional constraint to the covariance matrix of the array output vector. The regularization is obtained by subtracting a given amount of white noise from the main diagonal of the covariance matrix. Specifically, the DU beamformer aims at subtracting the signal subspace from the noisy signal space and it is computed by constraining the regularized covariance matrix to be negative definite. It is hence a data-dependent covariance matrix conditioning method. We show how to calculate precisely the unloading parameter, and we present an eigenvalue analysis for comparing the proposed DU beamforming, the minimum variance distortionless response (MVDR) filter and the multiple signal classification (MUSIC) method. Theoretical analysis and experiments with acoustic sources demonstrate that the DU beamformer localization performance is comparable to that of MVDR and MUSIC. Since the DU beamformer computational cost is comparable to that of a conventional beamformer, the proposed method can be attractive in array processing due to its simplicity, effectiveness and computational efficiency.
\end{abstract}

\begin{IEEEkeywords}
Diagonal unloading beamforming, source localization, direction of arrival estimation, broadband beamforming, acoustic analysis, array processing, microphone array.
\end{IEEEkeywords}

\IEEEpeerreviewmaketitle

\section{Introduction}

\IEEEPARstart{S}{ource} localization is an important task in array signal processing and it is hence of interest in different disciplines, including acoustics, communications, sonar, radar, astronomy, seismology, biomedicine. 

Beamforming is a robust method for source localization, which aims at estimating the source position by maximizing the power output of the spatial filter in the source direction. The conventional data-independent beamformer \cite{Bartlett1948} is based on a delay-and-sum procedure, which has its roots in time-series analysis. Without loss of generality, we consider herein narrowband and broadband beamforming in array processing applications for the direction of arrival (DOA) estimation problem assuming the source to be in the far-field. We examine the frequency-domain spatial filter and address the computation of the broadband beamformer by calculating the response power on each frequency bin and by fusing the narrowband components. 

The goal of a spatial filter is to leave undistorted the signal with a given DOA and to attenuate the response power for all the other directions. The minimum variance distortionless response (MVDR) \cite{Capon1969} beamformer is a well-know data-dependent filter which is aimed at minimizing the energy of noise and sources coming from different directions, while keeping a fixed gain on the desired DOA. It is based on the solution of an optimization problem that minimizes the power output subject to unity constraint in the look direction. However, its localization performance is not robust in most practical situations and, furthermore, the spatial spectrum might be deteriorated by steering vector errors and discrete sampling effects. Therefore, a robust variant of the MVDR filter, obtained with regularization techniques \cite{LiStoica2005}, is often preferred. In array processing, a popular approach to numerical stability improvement consists in acting on the conditioning of the correlation matrix before inverting it. This practice led to the class of diagonal loading (DL) techniques \cite{Cox1987,Carlson1988,GanzMosesWilson1990,LiStoicaWang2003, ElnoubiEl-Mikati2006, MestreLagunas2006, Chen2013}, which have been, and still are, deeply investigated by the scientific community due to their effectiveness. DL is implemented by imposing an additional quadratic constraint (penalty weight) to the objective function of the optimization problem that provides the optimal beamforming coefficients. In general, a limitation of the DL regularization is that it is not clear how to efficiently pick the penalty weight, although useful data-dependent methods have been proposed as \textit{ad hoc} procedures for specific applications \cite{MestreLagunas2006,LiuDing2008,DuLiStoica2010}. In practice, the regularization is obtained by adding an amount of white noise to the main diagonal of the covariance matrix. We will see that this operation is closely related to another popular localization method: the multiple signal classification (MUSIC) \cite{Schmidt1986}. MUSIC relies on the eigendecomposition of the covariance matrix and on the fact that the space spanned by its eigenvectors is partitioned into two orthogonal subspaces: the signal subspace and the noise subspace. MUSIC exploits the subspace orthogonality property to build the spatial spectrum and to localize the sources. The regularization in the MVDR beamformer can be interpreted as an operation that aims at emphasizing the orthogonality of the signal and noise subspaces. We demonstrate herein that, in the noise-free and single source case, with an appropriate choice of the penalty weight, the spatial spectrum of the MVDR is equal to the MUSIC spectrum, except for a scale factor. This property is illustrated and demonstrated here, and motivates the beamforming method proposed.

In this paper, we propose a regularized spatial filter based on the conventional beamforming. The regularization is data-dependent and it is obtained by subtracting an amount of white noise from the main diagonal of the covariance matrix of the array output vector. Although this diagonal unloading (DU) procedure is not new in the literature, as it has been used elsewhere to attenuate the effects of some noise components on the correlation matrix\cite{BrooksHumphreys1999,KimChoi2013}, the proposed DU beamforming is novel in the sense that it is based on two fundamental constraints: 1. the regularized covariance matrix has to be negative definite, and 2. the signal eigenvalue of the regularized covariance matrix has to be unique and equal to zero. Hence, we aim at removing as much as possible the signal subspace from the covariance matrix to design an high resolution beampattern. Note that in \cite{BrooksHumphreys1999,KimChoi2013}, the diagonal removal in the conventional beamforming is based on the fact that the spatially white noise is accumulated in the diagonal of the covariance matrix, and thus it can be removed with a subtraction operation on diagonal elements. The beneficial effect is to attenuate the noise in the beamforming computation without provoking any changes to the beampattern. In fact, in a noise-free scenario the beampattern is exactly that of the conventional spatial filter. Note instead that the proposed DU beamformer provides an high resolution beampattern. In our variant, the first constraint allows to put the proposed DU beamforming in a form which is comparable to the MVDR and MUSIC formulation. The second constraint determines exactly how to calculate the penalty weight for the DU operation. Note that for the theoretical derivation of the DU beamforming we will assume single source and noise-free conditions. Afterwards we will analyze the noisy case and the multisource scenario later on. The advantages of the DU beamforming are that the computation cost is comparable to that of the conventional beamforming whereas the localization performance and the spatial resolution is comparable to that of the regularized MVDR and MUSIC methods. 

In summary, the objective of this paper is twofold:
\begin{enumerate}
\item to derive a novel form of beamformer, which has high resolution and data-dependent properties, applying an appropriate diagonal unloading procedure to the conventional spatial filter and exploiting the subspace orthogonality property.
\item to provide an eigenvalue analysis highlighting the relationship between DU, MVDR and MUSIC by taking into account their regularized covariance matrices.
\end{enumerate}

The paper is organized as follows. Section \ref{sec1} provides the definition of the data model and of the narrowband and broadband spatial filters. The DU beamforming is then described in Section \ref{sec2}. We introduce the proposed beamformer by considering the noise-free and the single source case. Section \ref{sec3} provides an eigenvalue analysis of the covariance matrix and of the regularized covariance matrices in the DU beamformer, the MVDR filter with a DL regularization, and the MUSIC method. We provide some properties for evaluating the performance of the proposed method in comparison with high resolution beamforming. Next, an analysis of the DU beamforming in a noise and multisource scenario, and a computation cost comparison, are described in Section \ref{sec4}. Experiments using artificially-generated and real-world signals are shown in Section \ref{sec5}. The conclusions are drawn in Section \ref{sec6}.

\section{Background}
\label{sec1}

\subsection{Notation}

In this paper, we will make use of standard notational conventions. $\mathbb{R}$ and $\mathbb{C}$ denote the sets of all real and complex numbers respectively. Vectors
and matrices are written in boldface with matrices in capitals. For a random matrix $\mathbf{X}$, $E\{\mathbf{X}\}$ denotes the expectation of $\mathbf{X}$. For a matrix $\mathbf{A}$, $\mathbf{A}^T$, $\mathbf{A}^H$, and $\text{tr}(\mathbf{A})$ denote the transpose, the conjugate transpose, and the trace that is the sum of diagonal elements of $\mathbf{A}$, respectively. The identity matrix of any size are denoted by $\mathbf{I}$. The symbol * stands for convolution. 

\subsection{Data Model}

Suppose that a single source impinges upon an array of $N$ sensors and let $s(t) \in \mathbb{R}$ denotes the signal generated by a nonstationarity narrowband or broadband source at time $t$. 
The output of the n$th$ ($n=1,2,\dots,N$ ) sensor is given by
\begin{equation}
x_n(t)=h_n(t)*s(t-\tau_n)+v_n(t)
\end{equation}
where $h_n(t)$ is the impulse response from the source to the n$th$ sensor, $\tau_n$ is the propagation time from the source to the n$th$ sensor, $v_n(t)$ is an additive noise, which includes environmental background noise and electrical noise generated in the n$th$ channel. It is assumed to be uncorrelated and white Gaussian with zero mean and variance equal to $\sigma^2$ in all sensors. This 
is a reasonable model for many real-world noise fields \cite{KrimViberg1996}.
In the short-time Fourier transform domain, the $n$th received signal $X_n(k,f) \in \mathbb{C}$ for frequency bin $f$ is given by
\begin{equation}
X_n(k,f)=H_n(k,f) S(k,f) e^\frac{- j 2 \pi f \tau_n}{L}+V_n(k,f)
\end{equation}
where $k$ in the time block index, $X_n(k,f)$, $H_n(k,f)$, $S(k,f)$, and $V_n(k,f)$ are the discrete-time Fourier transforms (DTFTs) of $x_n(t)$, $h_n(t)$, $s(t)$, and $v_n(t)$ respectively, L is the size of the DTFT, and $j$ is the imaginary unit.
In vector notation, the data model of the array signals can be expressed as
\begin{equation}
\begin{split}
\mathbf{x}(k,f)&= \mathbf{E}(f)S(k,f)\mathbf{h}(k,f)+\mathbf{v}(k,f)\\
&=\mathbf{a}'(k,f,\theta)S(k,f)+\mathbf{v}(k,f)
\end{split}
\end{equation}
where
\begin{equation*}
\begin{split}
\mathbf{x}(k,f)&=[X_1(k,f), X_2(k,f), \dots, X_N(k,f)]^T \in \mathbb{C}^N,\\
\mathbf{h}(k,f)&=[H_1(k,f), H_2(k,f), \dots, H_N(k,f)]^T \in \mathbb{C}^N,\\
\mathbf{v}(k,f)&=[V_1(k,f), V_2(k,f), \dots, V_N(k,f)]^T \in \mathbb{C}^N,\\
\mathbf{E}(f)&=\text{diag}(e^\frac{- j 2 \pi f \tau_1}{L}, e^\frac{- j 2 \pi f \tau_2}{L}, \dots, e^\frac{- j 2 \pi f \tau_N}{L}) \in \mathbb{C}^{N \times N}
\end{split}
\end{equation*}
and $\mathbf{a}'(k,f,\theta) \in \mathbb{C}^N$ is the array steering vector of the source coming from direction $\theta$ defined as
\begin{equation}
\begin{split}
\mathbf{a}'(k,f,\theta)=&[H_1(k,f)e^\frac{- j 2 \pi f \tau_1}{L}, H_2(k,f)e^\frac{- j 2 \pi f \tau_2}{L}, \dots, \\
&\dots, H_N(k,f)e^\frac{- j 2 \pi f \tau_N}{L}]^T.\\
\end{split}
\label{asv}
\end{equation}
We now select the first sensor ($n=1$) as the reference sensor. Under the hypothesis that all the sensors are omnidirectional, identical, and have time-invariant transfer function, the expression (\ref{asv}) can be simplified in the far-field as \cite{StoicaMoses2005}
\begin{equation}
\mathbf{a}(f,\theta)=[1, e^\frac{- j 2 \pi f \tau_{12}(\theta)}{L}, \dots, e^\frac{- j 2 \pi f \tau_{1N}(\theta)}{L}]^T \in \mathbb{C}^N
\end{equation}
where $\tau_{1n}$ is the time difference of arrival (TDOA) between the reference sensor and sensor $n$. For a generic sensor $n$ paired with the reference sensor, the relationship between the TDOA $\tau_{1n}$ and the DOA $\theta$ is given by
\begin{equation}
\tau_{1n}(\theta)=\frac{d_{1n} \sin(\theta)}{c}
\label{doaeq}
\end{equation}
where $c$ is the speed of wave propagation and $d_{1n}$ is the distance between the reference and the $n$th sensor.

\subsection{Beamforming}

The output of a beamformer $Y(k,f,\theta) \in \mathbb{C}$ at time block $k$ for frequency $f$ in the look direction $\theta$ is obtained by weighting and summing the sensor signals
\begin{equation}
Y(k,f,\theta)=\mathbf{w}^H(k,f,\theta) \mathbf{x}(k,f)
\end{equation}
where $\mathbf{w}(k,f,\theta) \in \mathbb{C}^N$ is a vector for weighting and steering the data in the direction $\theta$. Then, the power spectral density (PSD) of the spatially filtered signal is
\begin{equation}
\begin{split}
P(k,f,\theta)&=E\{|Y(k,f,\theta)|^2\}\\
&=\mathbf{w}^H(k,f,\theta) \mathbf{\Phi} (k,f) \mathbf{w}(k,f,\theta)
\end{split}
\end{equation}
where $\mathbf{\Phi} (k,f)=E\{\mathbf{x}(k,f) \mathbf{x}^H(k,f)\} \in \mathbb{C}^{N \times N}$ is the PSD matrix of the convolved source signal, which is symmetric and positive definite. 
Let $P_s(k,f)=E\{S(k,f) S^H(k,f)\}$ denotes the power of the signal, then the PSD matrix of the array output vector can be written as
\begin{equation}
\mathbf{\Phi} (k,f)=P_s(k,f) \mathbf{a}(f,\theta) \mathbf{a}^H(f,\theta) + \sigma^2 \mathbf{I}.
\label{tcm}
\end{equation}
The spatially white noise is accumulated in the diagonal of the PSD matrix, and thus we can control the noise by properly modifying the diagonal elements.
The PSD matrix $\mathbf{\Phi}(k,f)$ is unknown and it has to be estimated from the received signal $\mathbf{x}(k,f)$ derived from the present and past signal blocks of the array
\begin{equation}
\widehat{\mathbf{\Phi}} (k,f)=\frac{1}{M}\sum_{k_p=0}^{M-1} \mathbf{x}(k-k_p,f)\mathbf{x}^H(k-k_p,f)
\label{ecm}
\end{equation}
where $M$ is the number of signal blocks for the averaging. Given the nonstationary nature of the source, we assume that the mean of the PSD matrix is computed in a short time in which the source can be considered stationary.

The PSD of a beamformer conveys information on the energy coming from direction $\theta$, and thus it should have a maximum peak in the direction of the source. Therefore, the localization estimation of the narrowband source is obtained by
\begin{equation}
\hat{\theta}=\underset{\theta}{\operatorname{argmax}} [P(k,f,\theta)].
\end{equation}
When the source is broadband, the PSD can be formalized by a parametric normalized incoherent frequency fusion \cite{Salvati2014} defined as
\begin{equation}
P(k,\theta)=\sum_{f=f_\text{min}}^{f_\text{max}}  \frac{P(k,f,\theta)}{(\underset{\theta}{\operatorname{max}}[P(k,f,\theta)])^\beta}
\label{pfn}
\end{equation}
where the parameter $\beta$ controls the level of normalization of the fusion process, and $f_\text{min}$ and $f_\text{max}$ denote the frequency range of the broadband source. When $\beta=1$ each narrowband component is normalized so that it has same weight in the fusion as the other components. This lends an high resolution to the spatial spectrum, but emphasizes the noise in those narrowband beamformers in which the signal-to-noise (SNR) ratio is low. When $\beta=0$ the fusion is not normalized. This provides poor resolution but possibly higher robustness against noise.

\section{Diagonal Unloading Beamforming}
\label{sec2}

We now formulate the DU beamforming. For simplicity we have omitted the dependence on time block $k$ in the rest of the paper. We make the following assumptions: 
\begin{enumerate}
\item We assume noise-free conditions and a single source scenario;
\item The weighting vector leaves undistorted the signals with a given DOA $\theta$;
\item The energy coming from all the other DOAs other than $\theta$ is attenuated as much as possible;
\item There exists a transformation $\mathcal{F}$ that turns the PSD matrix into a regularized PSD matrix $\mathbf{\Phi}_\text{DU} (f)=\mathcal{F}(\mathbf{\Phi} (f)) \in \mathbb{C}$ which is negative definite;
\item Only the eigenvalue $\lambda_s \in \mathbb{R}$ corresponding to the signal subspace of the PSD matrix $\mathbf{\Phi}$ has to be zero into the regularized PSD matrix $\mathbf{\Phi}_\text{DU}$. Hence, the eigenvalues $\lambda_v \in \mathbb{R}$ corresponding to the noise subspace of the regularized PSD matrix $\mathbf{\Phi}_\text{DU}$ have to be different from zero.
\end{enumerate}
We aim at totally remove the signal subspace from the PSD matrix using a DU regularization. This operation is the core of high resolution beamformers such as MVDR and MUSIC. In the MVDR, the attenuation of signal subspace is obtained by taking the inverse of the PSD matrix, while in the MUSIC the total removal is obtained by performing an eigendecomposition. We herein aim to remove the signal subspace by a diagonal unloading procedure. Therefore, the DU beamforming is given by the following optimization problem
\begin{equation}
\begin{split}
\text{minimize} \quad &\mathbf{w}^H(f,\theta) \mathbf{w}(f,\theta),\\
\text{subject to} \quad &\mathbf{w}^H(f,\theta) \mathbf{a}(f,\theta)=1,\\
	& \mathbf{w}^H(f,\theta) \mathbf{\Phi}_\text{DU} (f) \mathbf{w}(f,\theta)<0,\\
	& \lambda_s=0, \lambda_v\neq0 \quad \forall  \lambda_v.
\end{split}
\end{equation}
The solution is $\mathbf{w}=\mathbf{a}/N$, which is that of a conventional beamformer since only the unity constraint in the look direction affects the cost function.  By omitting the factor $1/N$ which has no influence on the DOA estimates, we can write the PSD as
\begin{equation}
P'_\text{DU}(f,\theta)=\mathbf{a}^H(f,\theta) \mathbf{\Phi}_\text{DU} (f) \mathbf{a}(f,\theta).
\label{du1}
\end{equation}
Since $\mathbf{\Phi}_\text{DU}$ is negative definite,  we can write the maximization PSD problem of equation (\ref{du1}) 
\begin{equation}
\hat{\theta}=\underset{\theta}{\operatorname{argmax}} [P'_\text{DU}(f,\theta)]
\end{equation}
in the following equivalent form
\begin{equation}
\hat{\theta}=\underset{\theta}{\operatorname{argmax}} [-\frac{1}{P'_\text{DU}(f,\theta)}]=\underset{\theta}{\operatorname{argmax}} [P_\text{DU}(f,\theta)]
\end{equation}
where the pseudo spatial spectrum $P_\text{DU}(f,\theta)$ is
\begin{equation}
P_\text{DU}(f,\theta)= -\frac{1}{\mathbf{a}^H(f,\theta) \mathbf{\Phi}_\text{DU} (f) \mathbf{a}(f,\theta)}.
\label{du2}
\end{equation}
To ensure that $\mathbf{\Phi}_\text{DU}$ is negative definite, each diagonal element has to be negative.  This operation can be computed by transforming the PSD matrix $\mathbf{\Phi}$ with a diagonal unloading. The DU regularized PSD matrix is given by
\begin{equation}
\mathbf{\Phi}_\text{DU} (f)=\mathbf{\Phi} (f) -\mu \mathbf{I}
\label{rdu}
\end{equation}
where $\mu$ is a real-valued, positive scalar.
Substituting equation (\ref{rdu}) in (\ref{du2}), the pseudo spatial spectrum becomes 
\begin{equation}
\begin{split}
P_\text{DU}(f,\theta)&=- \frac{1}{\mathbf{a}^H(f,\theta) (\mathbf{\Phi} (f) -\mu \mathbf{I}) \mathbf{a}(f,\theta)}\\
&= \frac{1}{\mathbf{a}^H(f,\theta) (\mu \mathbf{I} - \mathbf{\Phi} (f)) \mathbf{a}(f,\theta)}.
\label{duf}
\end{split}
\end{equation}
The PSD matrix $\mathbf{\Phi}$ can be decomposed in its eigenvalues and their associated eigenvectors through a subspace decomposition. Organizing the eigenvalues of $\mathbf{\Phi}$ in descending order ($\lambda_1 > \lambda_2 > \dots > \lambda_N, \lambda_n \in \mathbb{R}$) and denoting $\mathbf{u}_n \in \mathbb{C}^N$} their corresponding eigenvectors, the PSD matrix takes the following
form
\begin{equation}
\mathbf{\Phi} (f)=\mathbf{U} \mathbf{\Lambda}  \mathbf{U}^H
\end{equation}
where
\begin{equation*}
\begin{split}
\mathbf{\Lambda}&=
\begin{pmatrix}
\lambda_1\\
&\lambda_2\\
&&\ddots\\
&&&\lambda_N
\end{pmatrix},\\  
\mathbf{U}&=[\mathbf{u}_1, \mathbf{u}_2,\dots, \mathbf{u}_N].
\end{split}
\end{equation*}
Under the hypothesis of a single source, the eigenvector that correspond to the largest eigenvalue spans the signal subspace, and the remaining eigenvectors, which correspond to the smaller eigenvalues, span the noise subspace. Therefore, from (\ref{tcm}) we have that $\lambda_1=N P_s(f)+\sigma^2$, 
$\lambda_2=\lambda_3=\dots=\lambda_N=\sigma^2$, and that $\mathbf{u}_s=\mathbf{u}_1$ is the signal eigenvector. In fact, since each diagonal element of $\mathbf{\Phi} (f)$ in (\ref{tcm}) is equal to $P_s(f)+\sigma^2$, if $\sigma^2=0$ the signal eigenvalue is $\lambda_1=\text{tr}(\mathbf{\Phi} (f))=N P_s(f)$. In the noise-free case ($\sigma^2=0$) we can write the diagonal matrix of the eigenvalues as
\begin{equation}
\mathbf{\Lambda}=
\begin{pmatrix}
N P_s(f)\\
&0\\
&&\ddots\\
&&&0
\end{pmatrix}.
\end{equation}
The PSD matrix is thus singular and it can be written as
\begin{equation}
\mathbf{\Phi} (f)=N P_s(f) \mathbf{u}_s \mathbf{u}^H_s.
\end{equation}
By adding or subtracting a real quantity $\mu$ on each diagonal element of the PSD matrix, each eigenvalue is increased or decreased by the value $\mu$, the signal subspace remains the same and the noise subspaces are transformed since the diagonal modification can be interpreted as an injection or removal of white noise. In practice, a diagonal transformation of $\mathbf{\Phi}$ modifies the eigenvalues while keeping the proportions between eigenvectors. This fact is very important since we can control the contribution of the subspaces in the computation of the beamformer. Consider that the DL operation clearly guarantees the full-rank of the matrix to invert for the MVDR beamformer, but more meaningful the DL operation is closely related to the orthogonality property between signal and noise subspaces, which is the fundamental property on which the MUSIC method is built. These aspects will be analyzed in Section \ref{sec3}. 

The eigenvalue decomposition of the regularized PSD matrix (\ref{rdu}) is therefore given by
\begin{equation}
\mathbf{\Phi}_\text{DU} (f)=\bar{\mathbf{U}}
\begin{pmatrix}
 N P_s(f)-\mu\\
&-\mu\\
&&\ddots\\
&&&-\mu
\end{pmatrix}\bar{\mathbf{U}}^H
\end{equation}
where
\begin{equation}
\bar{\mathbf{U}}=[\mathbf{u}_s, \bar{\mathbf{u}}_2,\dots, \bar{\mathbf{u}}_N]
\end{equation}
and $\bar{\mathbf{u}}_v$, $v=2,\dots,N$ are the new eigenvectors of noise subspace.
Now, the constraint of having the eigenvalue corresponding to signal subspace of the regularized PSD matrix equal to zero becomes
\begin{equation}
 N P_s(f)-\mu=0.
\end{equation}
The solution is easily found by considering that the $\text{tr}(\mathbf{\Phi})=\text{tr}(\mathbf{\Lambda})= N P_s(f)$, and hence we have that the penalty weight of DU is data-dependent and is given by
\begin{equation}
\mu=\text{tr} (\mathbf{\Phi} (f)).
\label{pwf}
\end{equation}
The DU regularization with the penalty weight in (\ref{pwf}) guarantees that the regularized PSD matrix (\ref{rdu}) is negative definite and that the eigenvalues corresponding to noise subspace are non zero, since they are set to -$\mu$. This fact guarantees the total removal of signal subspace in the regularized PSD matrix. 
Finally, substituting (\ref{pwf}) in (\ref{duf}) the pseudo spatial spectrum of the DU beamforming is given by
\begin{equation}
P_\text{DU}(f,\theta)= \frac{1}{\mathbf{a}^H(f,\theta) (\text{tr} (\mathbf{\Phi} (f)) \mathbf{I} - \mathbf{\Phi} (f)) \mathbf{a}(f,\theta)}.
\label{dufinal}
\end{equation}
The DU regularized PSD matrix becomes
\begin{equation}
\mathbf{\Phi}'_\text{DU} (f)= \text{tr} (\mathbf{\Phi} (f)) \mathbf{I} - \mathbf{\Phi} (f).
\end{equation}
The constraint on the signal eigenvalue is fundamental for the proposed DU beamforming and has the important effect of improving the spatial resolution and the robustness against noise. In next section, the reasons of the constraint will be discussed by analyzing the eigendecomposition of the PSD matrix and of the regularized PSD matrix of the DU beamformer in comparison to the MVDR and MUSIC methods. 

\section{Eigeinanalysis of PSD Matrix}
\label{sec3}

We provide an analysis on the properties of the proposed DU beamforming. First, we briefly review the MVDR and the MUSIC methods. Then, we theoretically analyze the relationship between the beamformers by taking into account their eigenanalysis of their regularized PSD matrices. 

\subsection{The MVDR Beamformer}

The MVDR beamformer \cite{Capon1969} is a well-known data-dependent spatial filter technique which is aimed at minimizing the energy of noise and sources coming from different directions, while maintaining constant the gain on the desired direction. 
The MVDR filter using a DL regularization relies on the solution of the following minimization problem
\begin{equation}
\begin{split}
\text{minimize} \quad &\mathbf{w}^H(f,\theta)  (\mathbf{\Phi} (f)+\mu'  \mathbf{I}) \mathbf{w}(f,\theta),\\
\text{subject to} \quad &\mathbf{w}^H(f,\theta) \mathbf{a}(f,\theta)=1,\\
\end{split}
\label{mvdr}
\end{equation}
where $\mu'$ is a real-valued, positive scalar.
Solving (\ref{mvdr}) using the method of Lagrange multipliers, we obtain 
\begin{equation}
\mathbf{w}(f,\theta)=\frac{ (\mathbf{\Phi} (f)+\mu'  \mathbf{I})^{-1} \mathbf{a}(f,\theta)}{ \mathbf{a}^H(f,\theta)  (\mathbf{\Phi} (f)+\mu'  \mathbf{I})^{-1}  \mathbf{a}(f,\theta)}.
\end{equation}
Hence, the PSD of the regularized MVDR beamformer is given by
\begin{equation}
\begin{split}
P_\text{MVDR}(f,\theta)&= \frac{1}{\mathbf{a}^H(f,\theta) (\mathbf{\Phi} (f)+\mu'  \mathbf{I})^{-1} \mathbf{a}(f,\theta)}\\
&=\frac{1}{\mathbf{a}^H(f,\theta) \mathbf{\Phi}_\text{MVDR} (f) \mathbf{a}(f,\theta)}
\end{split}
\label{mvdrr}
\end{equation}
where the regularized PSD matrix of the MVDR is
\begin{equation}
\mathbf{\Phi}_\text{MVDR} (f)= (\mathbf{\Phi} (f)+\mu'  \mathbf{I})^{-1}.
\end{equation}
In the noise-free case, the diagonal matrix of the eigenvalues of $\mathbf{\Phi}_\text{MVDR}$ can be written
\begin{equation}
\mathbf{\Lambda}_\text{MVDR}(f)=
\begin{pmatrix}
\frac{1}{ N P_s(f)+\mu'}\\
&\frac{1}{\mu'}\\
&&\ddots\\
&&&\frac{1}{\mu'}
\end{pmatrix}.
\end{equation}
Since $\mathbf{\Phi}_\text{MVDR}$ is Hermitian and full-rank, the eigenvectors of the inversion matrix are the same as the eigenvectors of the matrix. Thus, the eigenvector of the signal subspace remains the same, and the regularization transforms only the noise subspace.
Basically, the DL regularization aims at reducing the eigenvalue of the signal subspace, and returns larger eigenvalues corresponding to the noise subspace, since the penalty weight $\mu'$ has in general a small value.

\subsection{The MUSIC Beamformer}

The MUSIC beamformer \cite{Schmidt1986} is based on an eigendecomposition which exploits the orthogonality between signal and noise subspaces. The estimated noise subspace is used for obtaining the steering vector that is as orthogonal to the noise subspace as possible. The subspace orthogonality property leads us to define the pseudo spatial spectrum
\begin{equation}
P_\text{MUSIC}(f,\theta)=\frac{1}{\mathbf{a}^H(f,\theta) \mathbf{U}_v(f) \mathbf{U}_v^H(f)\mathbf{a}(f,\theta)}
\label{music}
\end{equation}
where $\mathbf{U}_v(f) \in \mathbb{C}^{N \times (N-1)}$ is a matrix containing the eigenvectors corresponding to the noise subspace
\begin{equation}
\mathbf{U}_v(f)=[\mathbf{u}_2, \mathbf{u}_3,\dots, \mathbf{u}_N].
\end{equation}
Equation (\ref{music}) can be written as
\begin{equation}
P_\text{MUSIC}(f,\theta)=\frac{1}{\mathbf{a}^H(f,\theta) \mathbf{\Phi}_\text{MUSIC}(f) \mathbf{a}(f,\theta)}
\label{mus}
\end{equation}
where the regularized PSD matrix $\mathbf{\Phi}_\text{MUSIC}$ is expressed as
\begin{equation}
\mathbf{\Phi}_\text{MUSIC}(f)=\mathbf{U}(f) \Lambda_\text{MUSIC}(f) \mathbf{U}^H(f).
\label{musicr}
\end{equation}
We can interpret the MUSIC as a beamformer that uses a regularized PSD matrix $\mathbf{\Phi}_\text{MUSIC}$ in which the eigenvector matrix $\mathbf{U}$ is that of $\mathbf{\Phi}$ and the eigenvalue diagonal matrix is
\begin{equation}
\mathbf{\Lambda}_\text{MUSIC}(f)=
\begin{pmatrix}
0\\
&1\\
&&\ddots\\
&&&1
\end{pmatrix}.
\end{equation}
Therefore, MUSIC assigns a zero value for the eigenvalue corresponding to the signal subspace and value 1 for each eigenvalue corresponding to the noise subspace.

\subsection{Eigenvalues Analysis Comparison}

We now consider the eigenvalue analysis for comparing the proposed DU beamforming with MVDR and MUSIC. By writing MVDR and MUSIC in form of regularized PSD matrix function of the PSD matrix
we have the same form for the spatial spectrum of the two beamformers (\ref{mvdrr}) and (\ref{mus}).

\textit{Theorem 1}: Suppose a single source and $\sigma^2=0$, then there exists a penalty weight $\mu' \in \mathbb{R}$ for the MVDR such that $P_\text{MVDR}\approx k_1 P_\text{MUSIC}$, $k_1 \in \mathbb{R}^+$.

\textit{Proof}: We have
\begin{equation}
\mathbf{\Phi}_\text{MVDR} (f)\approx \frac{1}{k_1} \mathbf{\Phi}_\text{MUSIC}(f) 
\label{eqm}
\end{equation}
and we can write  equation (\ref{eqm}) in a form which is proportional to the  eigenvalue matrix
\begin{equation}
\mathbf{\Lambda}_\text{MVDR} (f)\approx k_2 \mathbf{\Lambda}_\text{MUSIC}(f) 
\end{equation}
where $k_2$ is a real positive value. We can note that multiplying the eigenvalue matrix $ \mathbf{\Lambda}_\text{MUSIC}$ with a constant $k_2$ has the only effect of scaling the spatial spectrum. Specifically, we can write the normalization eigenvalue matrix of the MVDR by multiplying each element for $\mu'$ and we obtain
\begin{equation}
\bar{\mathbf{\Lambda}}_\text{MVDR}(f)=
\begin{pmatrix}
\frac{\mu'}{N P_s(f)+\mu'}\\
&1\\
&&\ddots\\
&&&1
\end{pmatrix}.
\end{equation}
The equality $P_\text{MVDR}=k_1 P_\text{MUSIC}$ is theoretically achieved when $P_s(f)=\infty$. The approximation is demonstrated when $\mu'$ has a very small value in comparison with $N P_s(f)$, and we can consider $\frac{\mu'}{N P_s(f)+\mu'}$ to be nearly zero. On the other hand, a large value of $P_s(f)$ or $N$ makes the eigenvalue smaller.

Theorem 1 shows that the regularization in the MVDR is an operation that exploits the orthogonality property of the signal and noise subspaces. In general, for the noise-free case the best regularization for the MVDR is thus equal to the implementation of the MUSIC method. It is interesting to note the effect of considering the signal subspace in the MUSIC by changing the eigenvalue of the signal subspace in the range $0\leq \lambda_1\leq 1$. When $\lambda_1$ tends to 1 the searching procedure of the orthogonality is degraded due to the attenuation of the energy coming from all the other DOAs different from the look direction $\theta$. 

The DU beamformer in (\ref{dufinal}) has also a similar form if compared to the MVDR and MUSIC beamformers in (\ref{mvdrr}) and (\ref{mus}).

\textit{Theorem 2}: Suppose a single source and and $\sigma^2=0$, then $P_\text{DU}= k_3 P_\text{MUSIC}$, $k_3 \in \mathbb{R}^+$.

\textit{Proof}: We can write the equality $P_\text{DU}= k_3 P_\text{MUSIC}$ by considering the eigenvalue matrices
\begin{equation}
\mathbf{\Lambda}_\text{DU} (f)= k_4 \mathbf{\Lambda}_\text{MUSIC}(f)
\end{equation}
where $k_4$ is a real positive value. From (\ref{dufinal}) we have that the regularized matrix for the DU beamforming can be written as $\mathbf{\Phi}'_\text{DU} (f)=(\text{tr} (\mathbf{\Phi} (f)) \mathbf{I} - \mathbf{\Phi} (f))$, and therefore its eigenvalue matrix is given by
\begin{equation}
\mathbf{\Lambda}_\text{DU}(f)=
\begin{pmatrix}
0\\
&\text{tr} (\mathbf{\Phi} (f)) \\
&&\ddots\\
&&&\text{tr} (\mathbf{\Phi} (f)) 
\end{pmatrix}.
\end{equation}

In this case, since the signal eigenvalue of the DU regularized PSD matrix is zero, we have that DU differs from MUSIC only for a scaling factor of the spatial spectrum, and we have $k_4=\text{tr} (\mathbf{\Phi} (f)) $. Besides that, the DU beamforming has two advantages in the noise-free case if compared to MVDR. First, DU guarantees that the signal subspace is not used in the orthogonality searching procedure, while the MVDR guarantees this condition only in the ideal case $P_s(f)=\infty$, although a small contribution may be a negligible factor in the localization performance. Second, DU does not require any choice of the regularization parameter, while for the MVDR using the DL, a penalty weight $\mu'$ has to be found empirically for an optimal performance. We stress the fact that Theorem 1 and Theorem 2 are valid when only one source impinges the array in the noise-free case. In the next section, we provide an analysis referred to a more realistic noise scenario, and in a multisource case. 

\section{Analysis of the DU beamforming}
\label{sec4}

In this section, we derive the DU beamforming in a noise scenario by considering the spatially white noise $\sigma^2$. Then, we analyze the multisource case and examine the computational cost. 

\subsection{Noisy Environment Scenario}

We have introduced in Section \ref{sec2} the DU beamforming with the spatially white noise $\sigma^2=0$ for a better understanding of the diagonal removal procedure and of the relationship with MVDR and MUSIC. In particular, we have seen the importance of having only the signal eigenvalue equal to zero, which also implies that the DU regularized PSD matrix is negative definite.

However, a more realistic condition is related to a noisy environment scenario, and we can therefore write the eigenvalue matrix of the PSD matrix $\mathbf{\Phi}$ as
\begin{equation}
\mathbf{\Lambda}=
\begin{pmatrix}
N P_s(f)+\sigma^2\\
&\sigma^2\\
&&\ddots\\
&&&\sigma^2
\end{pmatrix}.
\end{equation}
The trace of $\mathbf{\Phi}$ becomes
\begin{equation}
\text{tr} (\mathbf{\Phi} (f))=\text{tr} (\mathbf{\Lambda})=N(P_s(f)+\sigma^2).
\end{equation}
The DU in equation (\ref{dufinal}) does not guarantees now that the signal eigenvalue of the regularized PSD matrix is zero. We have that the signal eigenvalue $\lambda_1$ after the diagonal removing is
\begin{equation}
\lambda_1=N(P_s(f)+\sigma^2)-N P_s(f)+\sigma^2=(N-1)\sigma^2.
\end{equation}
In noisy conditions, the pseudo spatial spectrum of the DU beamforming becomes
\begin{equation}
P_\text{DU}(f,\theta)= \frac{1}{\mathbf{a}^H(f,\theta) \mathbf{\Phi}''_\text{DU} \mathbf{a}(f,\theta)}.
\end{equation}
where
\begin{equation}
\mathbf{\Phi}''_\text{DU}=[\text{tr} (\mathbf{\Phi} (f))-(N-1)\sigma^2] \mathbf{I} - \mathbf{\Phi} (f).
\end{equation}
If $\sigma^2$ is known, the optimal implementation of the DU beamforming is obtained. Typically $\sigma^2$ can be estimated from a few signal-free analysis blocks, if the noise can be considered stationary.
Otherwise, the DU implementation is not optimal, Theorem 2 is not valid, and the pseudo spatial spectrum can be represented only by an approximation. In such case, a certain quantity of the signal eigenvector is used in the beamforming computation and may result in some degradation in the localization performance due to a reduced effectiveness in exploiting the orthogonality of subspaces. Let $G$ and SNR$=P_s(f)/\sigma^2$ denote the gain of the signal subspace in the spatial filter and the signal-to-noise ratio respectively. We can then write the gain as
\begin{equation}
G=\frac{(N-1)\sigma^2}{N(P_s(f)+\sigma^2)}=\frac{(N-1)}{N(\text{SNR}+1)}.
\end{equation}
The gain is in general low, and it can be a negligible factor in many cases such as in MVDR.

\subsection{Multisource Scenario}

Without loss of generality, we consider the case in which two sources impinge an array of sensors. Let $s_1(t) \in \mathbb{R}$ and $s_2(t) \in \mathbb{R}$ denote the signals generated by two sources at time $t$. We assume that the sources can be both narrowband or both broadband. The PSD matrix can be written as
\begin{equation}
\mathbf{\Phi} (f)=\mathbf{A}\mathbf{S}\mathbf{A}^H + \sigma^2 \mathbf{I}
\label{rmus}
\end{equation}
where
\begin{equation*}
\begin{split}
\mathbf{A}&=[\mathbf{a}_1(f,\theta),\mathbf{a}_2(f,\theta)],\\ 
\mathbf{S}&=
\begin{pmatrix}
P_{s_1}(f) & 0\\
0 &P_{s_2}(t,f)\\
\end{pmatrix},
\end{split}
\end{equation*}
with $\mathbf{a}_1(f,\theta)$ and $\mathbf{a}_2(f,\theta)$ being the array steering vectors for source $s_1(t)$ and $s_2(t)$, and $P_{s_1}$ and $P_{s_2}$ being the power of the source signals.
Let $\lambda_1$ and $\lambda_2$ denote the larger eigenvalues corresponding to the signal subspaces of $s_1(t)$ and $s_2(t)$, then we have
\begin{equation}
\begin{split}
\lambda_1&=N P_{s_1}(f)+\sigma^2,\\
\lambda_2&=N P_{s_2}(f)+\sigma^2,\\
\lambda_v&=\sigma^2, \quad v=3,\dots,N.
\end{split}
\end{equation}
With a DU operation it is impossible to reduce the eigenvalues of the signal subspaces to zero. In a multisource scenario, we can only minimizing these values. Let $G_1$ and $G_2$ denote the gains of the signal subspace in the regularized PSD matrix, then we have
\begin{equation}
\begin{split}
G_1=\frac{NP_{s_2}(f)+(N-1)\sigma^2}{N(P_{s_1}(f)+P_{s_2}(f)+\sigma^2)},\\
G_2=\frac{N(P_{s_1}(f)+(N-1)\sigma^2}{N(P_{s_1}(f)+P_{s_2}(f)+\sigma^2)}.
\end{split}
\end{equation}
Considering SNR$_1=P_{s_1}(f)/\sigma^2$ and SNR$_2=P_{s_2}(f)/\sigma^2$, we have
\begin{equation}
\begin{split}
G_1=\frac{N \text{SNR}_2+(N-1)}{N(\text{SNR}_1+\text{SNR}_2+1)},\\
G_2=\frac{N \text{SNR}_1+(N-1)}{N(\text{SNR}_1+\text{SNR}_2+1)}.
\end{split}
\end{equation}
The attenuation of a signal subspace is thus related to the SNR of both sources. In the general case of $S$ sources, each signal subspace vector is attenuated by a factor proportional to the corresponding signal eigenvalue and the gain quantity is related to the sum of the other eigenvalues due to the subtraction operation.

\subsection{Computational Cost}

The computational cost of the DU beamformer is that of a conventional spatial filter. In fact, the diagonal removing is a negligible operation since it consists of additions or subtractions. Besides that, the inversion due to the pseudo spatial spectrum at the denominator is ininfluent, since we can consider only the spatial spectrum at the numerator and we can search it for the minimum value to estimate the source position. On the other hand, the MVDR and MUSIC methods require a full-rank inversion matrix the former, and an eigendecomposition the latter. For both methods, there is the need of a singular value decomposition which has complexity $O(N^3)$.
Hence, the DU beamforming is attractive in array processing since it provides an higher resolution at no additional cost.

In table \ref{allpsd}, we summarize the narrowband spatial spectrum equations of the proposed DU beamforming, the MVDR, and the MUSIC omitting for simplicity the dependency from $f$, and $\theta$. We consider the non-optimal implementation of the DU procedure assuming unknown the noise $\sigma^2$.

\begin{table}[!t]
\centering
\renewcommand{\arraystretch}{2.0}
\caption{The spatial spectrum of DU, MVDR and MUSIC beamformers.}
\label{allpsd}
\begin{tabular}{@{}ll@{}}
\toprule
\midrule
DU & $P_\text{DU}= \frac{1}{\mathbf{a}^H (\text{tr} (\mathbf{\Phi} ) \mathbf{I} - \mathbf{\Phi} ) \mathbf{a}}$ \\
MVDR & $P_\text{MVDR}= \frac{1}{\mathbf{a}^H (\mathbf{\Phi} +\mu'  \mathbf{I})^{-1} \mathbf{a}}$\\
MUSIC & $P_\text{MUSIC}=\frac{1}{\mathbf{a}^H \mathbf{U}_v \mathbf{U}_v^H\mathbf{a}}$\\
\midrule
\bottomrule
\end{tabular}
\end{table}

\section{Experimental Results}
\label{sec5}

In this section, we present some numerical and real-world results to validate the proposed DU beamformer. We compare the DOA estimation performance with state of the art methods in the context of acoustic source localization using a microphone array. 

\subsection{Synthetic Data}

We have considered an uniform linear array in free-field and reverberant conditions, and have performed acoustic simulations with acoustic sources modeled as nonstationary broadband speech signal and stationary broadband USASI noise. We have investigated both the single and the multiple source scenarios. The sources and microphones were considered omnidirectional. We have compared the DOA localization performance using the root mean square error (RMSE) of the proposed DU, the MVDR \cite{Capon1969} with DL, the MUSIC \cite{Schmidt1986}, the conventional beamforming, i.e, the steered response power (SRP), and the SRP phase transform (SRP-PHAT) \cite{DiBiase2001}. The latter is a conventional beamformer in which the narrowband components are first normalized by taking into account only the phase information, and then fused to obtain the broadband beamformer. Note that in all other methods the broadband fusion was instead computed with the post-filter normalization \cite{Salvati2014} of equation (\ref{pfn}). We have assumed $\beta=1$. Hence, we have considered the same weight of each narrowband component in the fusion. Note that this fact involves an high resolution in the broadband localization problem \cite{Salvati2014}. A data-dependent DL for the MVDR is adopted to improve the robustness of the MVDR.
The data-dependent DL factor used in these simulations is given by
\begin{equation}
\mu'(f)=\frac{1}{L}\text{tr} (\mathbf{\Phi} (f)) \Delta
\end{equation}
where $\Delta$ is the loading constant and $L$ is the size of the DTFT. We have set $\Delta$ equal to $10^{-4}$ since a small value keeps an high resolution in each narrowband beamformer. Note that an increasing of $\Delta$ may result in a greater signal eigenvalue in the regularized PSD matrix. This fact implies that a larger amount of signal subspace is used in the PSD estimation, reducing the resolution of the MVDR.

\begin{figure}[t]
\centering
\includegraphics[width=1.0\columnwidth]{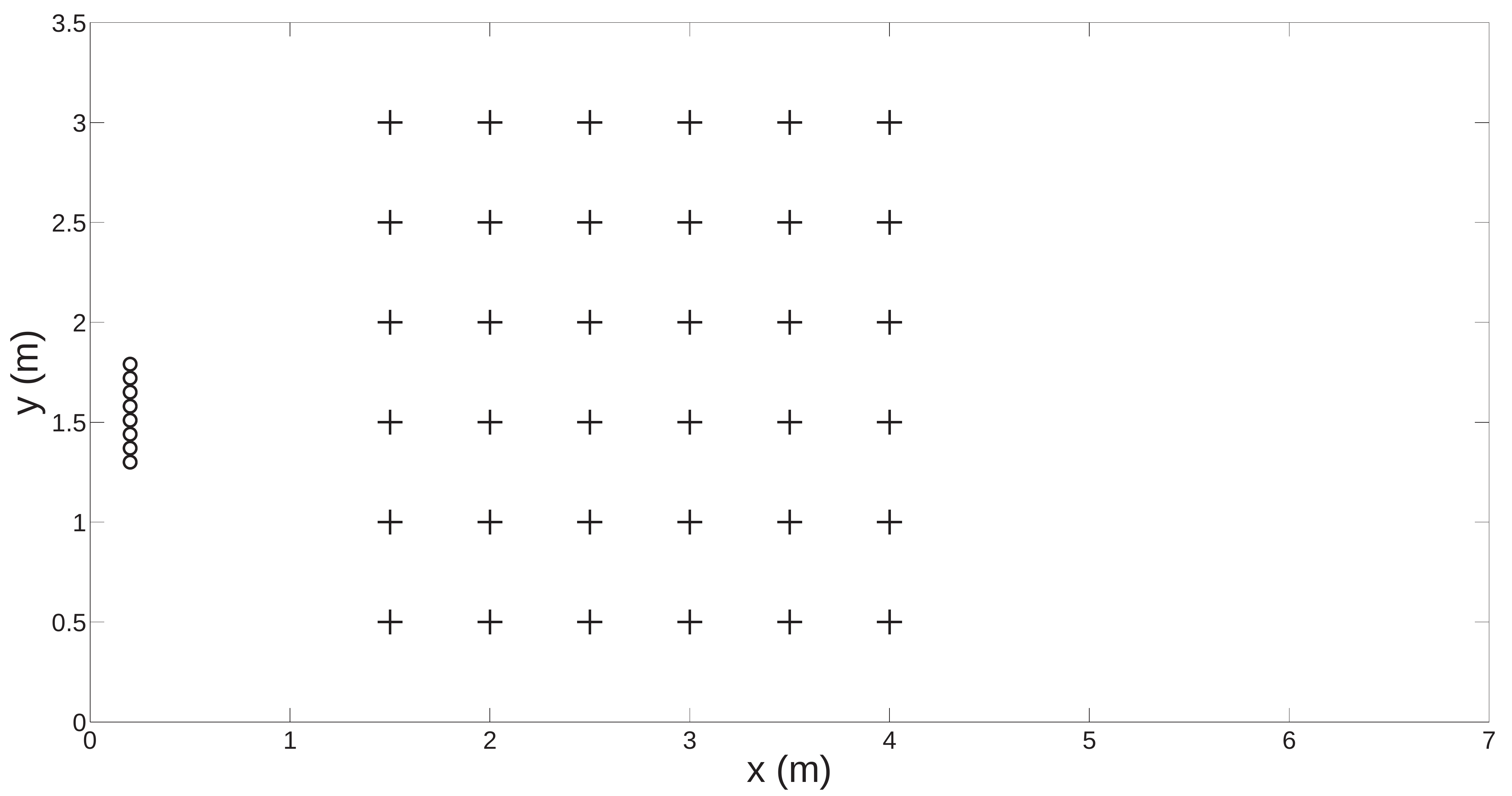}
\caption{The simulated room setup with the positions of sensors and sources.}
\label{ss_setup}
\end{figure}

The inversion and eigendecomposition of the PSD matrix was performed by single value decomposition since it provides some numerical advantages. Besides, an optimal frequency range between 80 Hz and 8000 Hz, since it is the typical spectrum range of speech signals, was used for all beamformers. The sampling frequency was 44.1 kHz and the window size L was 2048 samples with an overlap of 512 samples. Given the nonstationary nature of the speech signal, a small number of snapshots has been considered for the estimation of the PSD matrix. The simulations were conducted with different SNR levels, obtained by adding mutually independent white Gaussian noise to each channel.
For the DU beamforming, we have assumed that noise is unknown and we thus used equation (\ref{dufinal}) for the computation of the pseudo spatial spectrum. 

\begin{figure}[t]
\centering
\includegraphics[width=1.0\columnwidth]{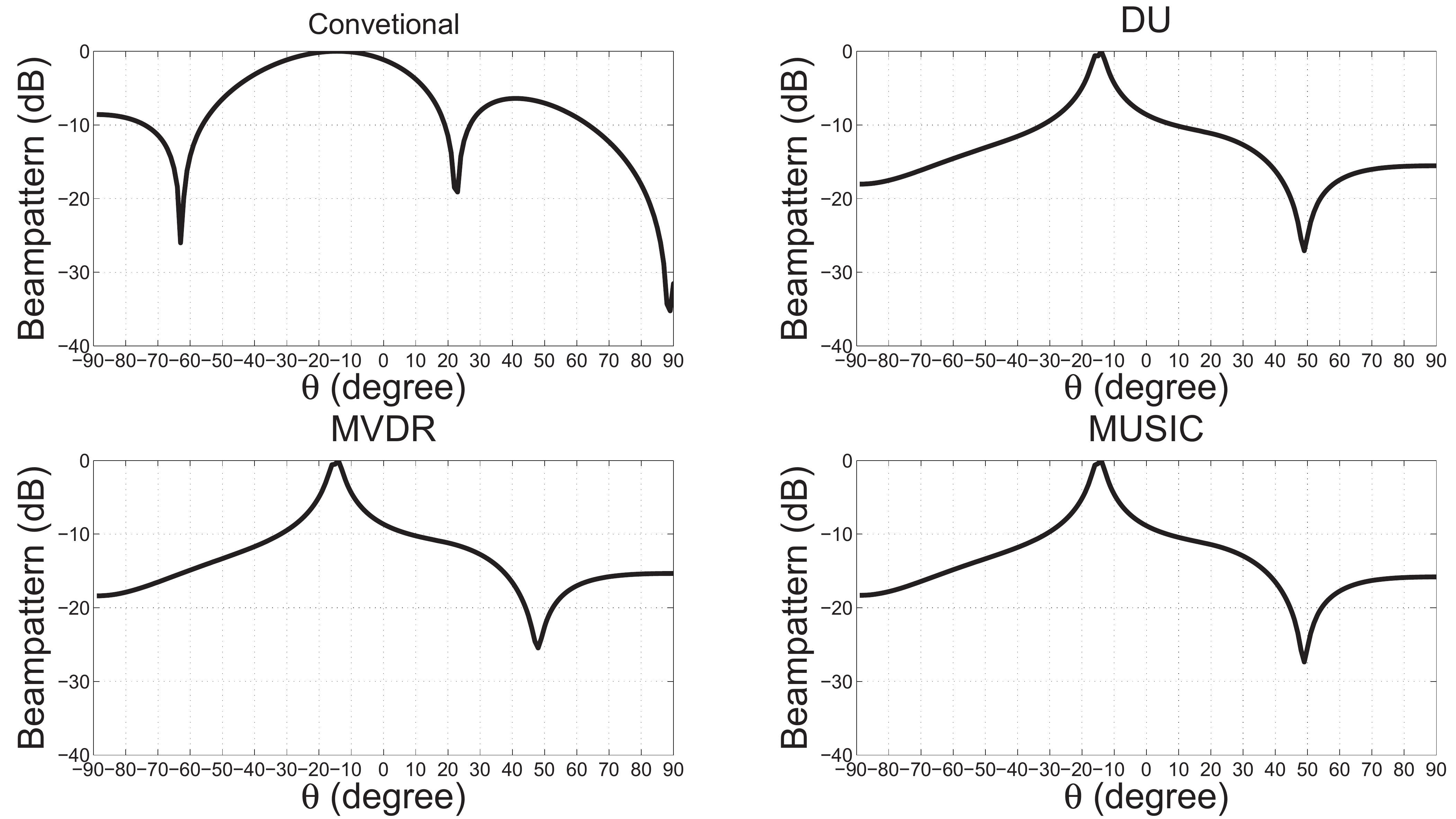}
\caption{The beampattern for the conventional, the DU, the MVDR, and the MUSIC beamformers of a 1000 Hz sinusoid impinging the array from a DOA of -18 degree.}
\label{bp_8m}
\end{figure}

\subsection{The Single Source Case}

In the experiments with single source simulated data, an uniform linear array of 8 sensors was used. We have investigated the free-field case and the reverberant case. Acoustic simulations of reverberant environments were obtained with the image-source method \cite{Allen1979}. A localization task in a room of 7 m $\times$ 3.5 m $\times$ 3 m was considered. The distance between microphones was 0.07 m. The room setup is shown in Figure \ref{ss_setup} in which we can see the source positions used in the simulations. 

First, we analyzed the narrowband responses when a sinusoidal signal impinges the array. The beampatterns for a frequency of 1000 Hz with a DOA of -18 degree is shown in Figure \ref{bp_8m}. 
The beampattern is calculated by considering that the weighting steering vector of the MVDR beamformer is
\begin{equation}
\mathbf{w}_\text{MVDR}(f,\theta)=\frac{ \mathbf{\Phi}_\text{MVDR} (f) \mathbf{a}(f,\theta)}{ \mathbf{a}^H(f,\theta)  \mathbf{\Phi}_\text{MVDR} (f)  \mathbf{a}(f,\theta)}.
\end{equation}
Since the similar forms of spatial spectrum as reported in Table \ref{allpsd}, the weighting steering vectors of DU and MUSIC can be written as
\begin{equation}
\mathbf{w}_\text{DU}(f,\theta)=\frac{ \mathbf{\Phi}'_\text{DU} (f) \mathbf{a}(f,\theta)}{ \mathbf{a}^H(f,\theta)  \mathbf{\Phi}'_\text{DU} (f)  \mathbf{a}(f,\theta)},
\end{equation}
\begin{equation}
\mathbf{w}_\text{MUSIC}(f,\theta)=\frac{ \mathbf{\Phi}_\text{MUSIC} (f) \mathbf{a}(f,\theta)}{ \mathbf{a}^H(f,\theta)  \mathbf{\Phi}_\text{MUSIC} (f)  \mathbf{a}(f,\theta)}.
\end{equation}
We can observe the high resolution response of the DU beamforming. 

\begin{figure}[t]
\centering
\includegraphics[width=1.0\columnwidth]{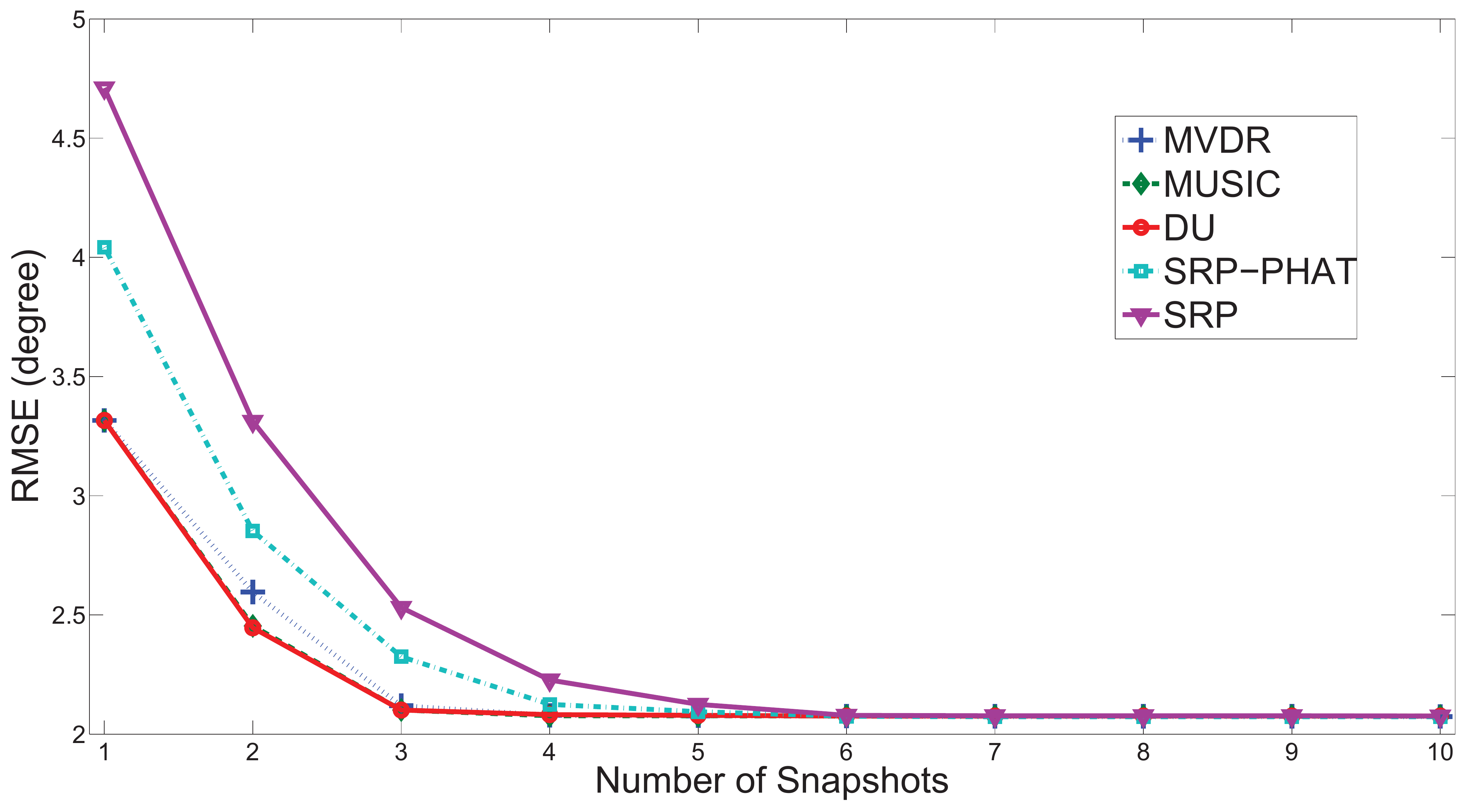}
\caption{Localization performance of a male speech signal in free-field at variation of number of snapshots with a SNR of 20 dB.}
\label{ss_snap}
\end{figure}

\begin{figure}[t]
\centering
\includegraphics[width=1.0\columnwidth]{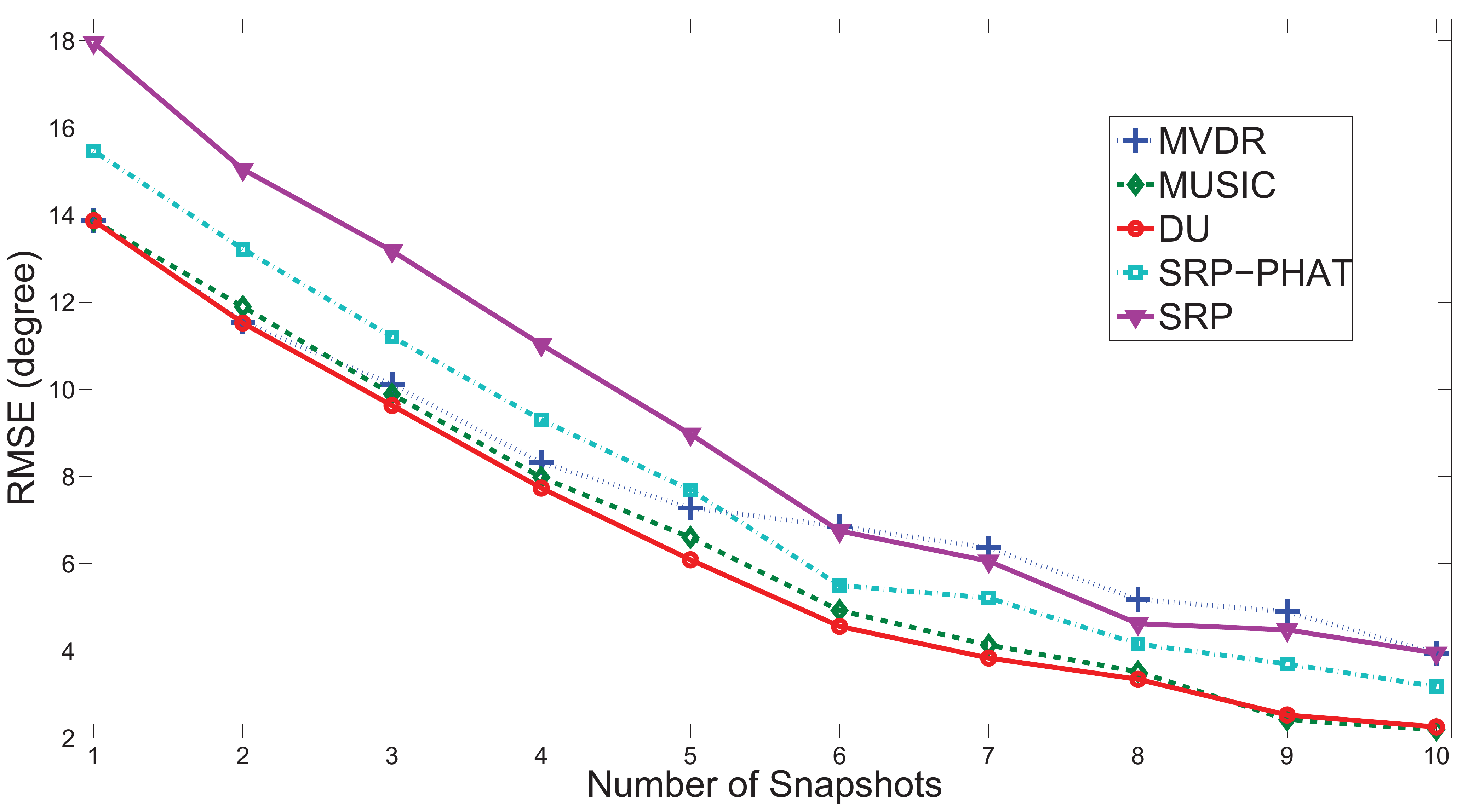}
\caption{Localization performance of a male speech signal in free-field at variation of number of snapshots with a SNR of 0 dB.}
\label{ss_snap_zero}
\end{figure}

Next, the localization of a male speech signal in free-field condition was investigated. The DOA estimation performance for different number of snapshots used to estimate the PSD matrix was conducted. Figures \ref{ss_snap} and \ref{ss_snap_zero} show the results for a SNR of 20 dB and 0 dB respectively. The DU performance is in general superior or comparable to the other considered methods. Note that in the case of a single snapshot, the localization performance is the same for DU, MVDR and MUSIC since the estimated PSD matrix provides only one large eigenvalue for the signal subspace and the noise eigenvalues are null. When the number of snapshots increases, all beamformers tend to the best performance that can be reached for this system in case of moderate noise (Figure \ref{ss_snap}). When the noise is higher (Figure \ref{ss_snap_zero}), DU and MUSIC show the best performance while the MVDR tends to degrade when the number of snapshots increases above 5. In fact, the PSD matrix needs of a large number of snapshots to be accurately estimated. In case of few snapshots and a low SNR, the noise eigenvalues may have different and close values, resulting in a distinct weighting of the signal subspace for DU and MVDR. The inversion operation tends to provide a greater difference in the output for small differences in the eigenvalues of the regularized PSD matrix. This fact is the reason of the worse performance in higher noise conditions for the MVDR. Note that \textit{ad hoc} regularization $\mu'$ can improve the performance by minimizing the differences on noise eigenvalues.  On the other hand, the DU is based on a subtraction operation resulting in a less marked difference of noise subspace eigenvalues. Specifically, in some cases the DU beamforming outperforms the MUSIC method. As we observe in Figure \ref{ss_snap} for the moderate noise case, when the number of snapshots is 2, DU results in a smaller RMSE error if compared to that of MUSIC. For the high noise case depicted in Figure \ref{ss_snap_zero}, we can observe a better performance of DU in the range [2-8] of the snapshot number. This fact can be interpreted as a different weighting of noise subspaces in the DU, which may result in a set of less relevant noise eigenvectors, corresponding to the smaller eigenvalues of the PSD matrix. Hence, DU proves to be robust with respect to PSD estimation errors. Note that MUSIC assumes that each noise eigenvalue is equal to $1$, resulting in the same weighting for all the noise eigenvectors. 

\begin{figure}[t]
\centering
\includegraphics[width=1.0\columnwidth]{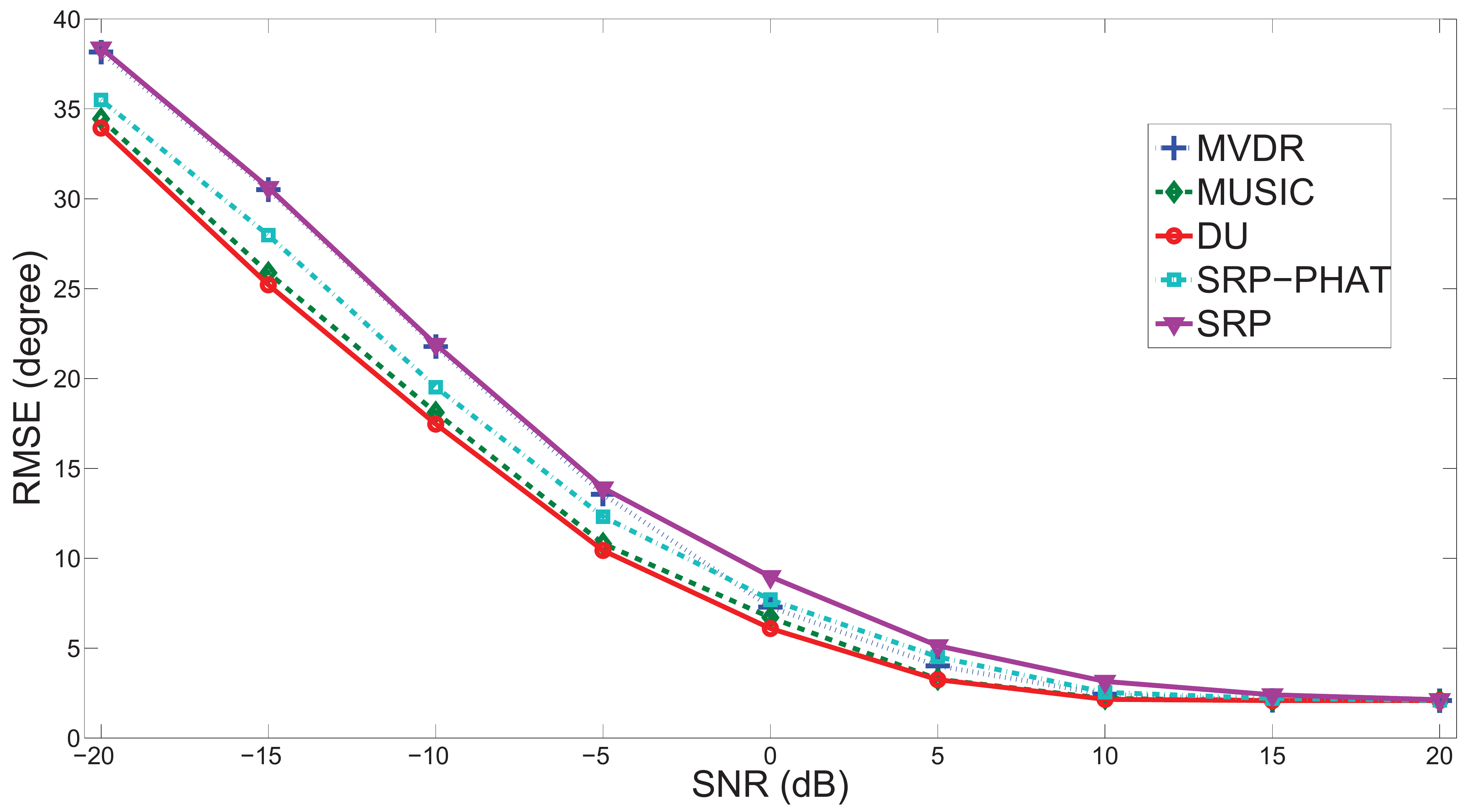}
\caption{Localization performance of a male speech signal in free-field at variation of SNR with a number of snapshots of 5.}
\label{ss_noise}
\end{figure}

Finally, we have conducted an analysis for different values of SNR in the range [-20,20] dB, with a number of snapshots equal to 5. Figure \ref{ss_noise} shows the results. DU shows the best performance, which is very close to that of MUSIC, whereas the MVDR degrades when the SNR decreases for the reasons mentioned above.

\begin{figure}[t]
\centering
\includegraphics[width=1.0\columnwidth]{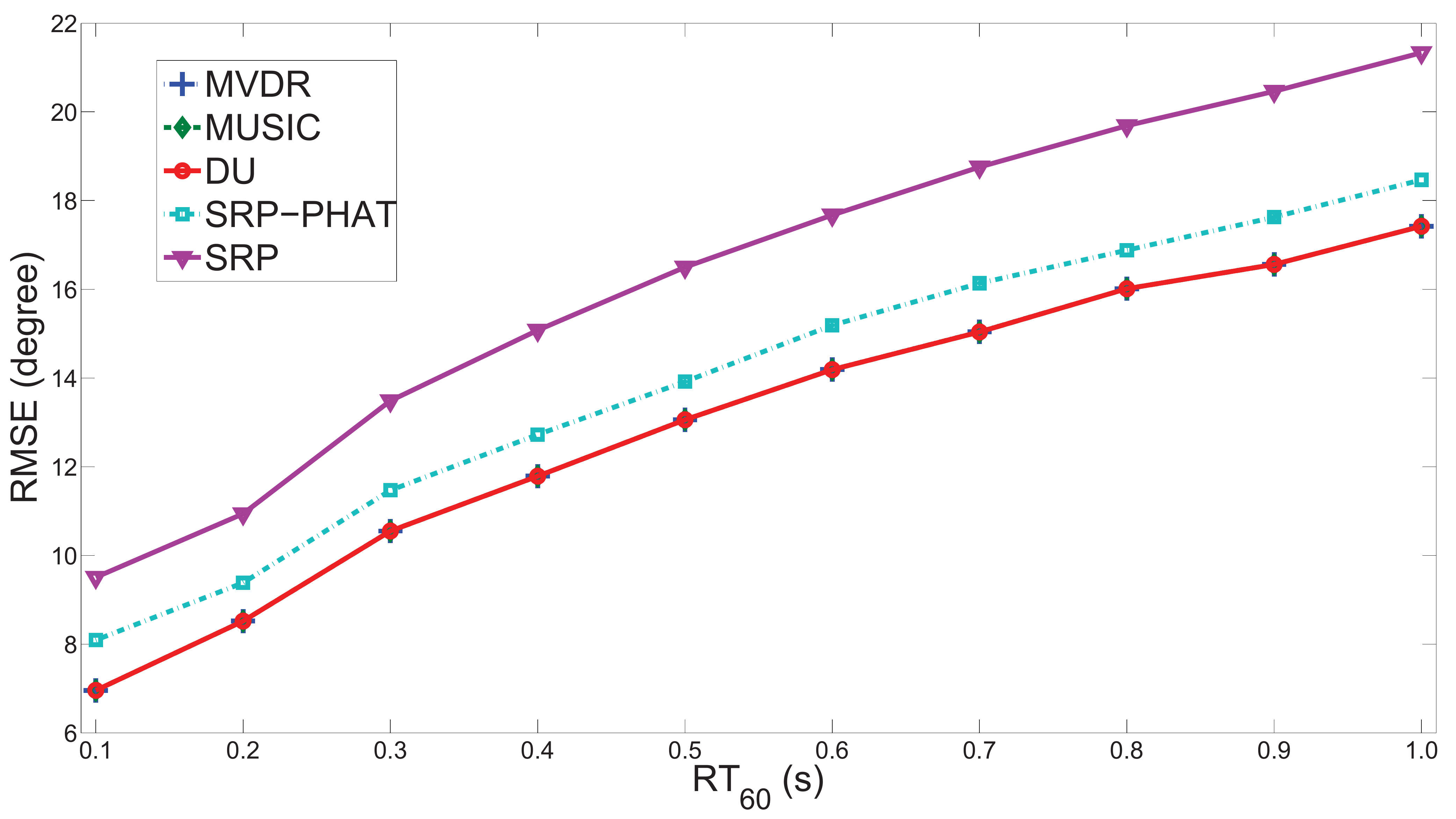}
\caption{Localization performance of a male speech signal at variation of RT$_{60}$ with a SNR of 10 dB and a single snapshot.}
\label{ss_rt_1s}
\end{figure}

\begin{figure}[t]
\centering
\includegraphics[width=1.0\columnwidth]{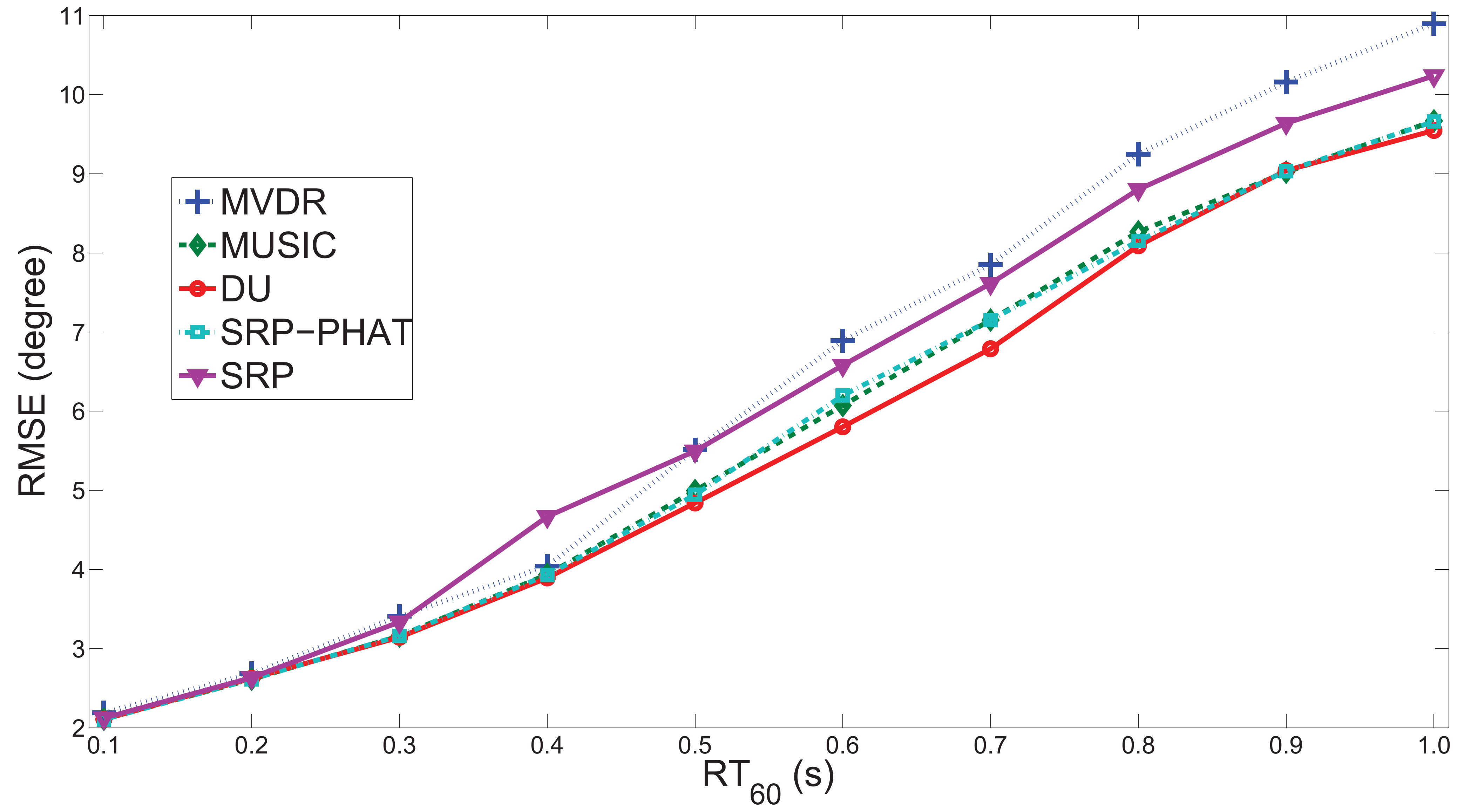}
\caption{Localization performance of a male speech signal at variation of RT$_{60}$ with a SNR of 10 dB and a number of snapshots of 10.}
\label{ss_rt_10s}
\end{figure}

Next, simulations were conducted to analyze the effects of reverberation on the DU algorithm. We have examined two numerical examples for the localization performance at variation of reverberation time (RT$_{60}$) with a single snapshot and a number of snapshots equal to 10. As it can be seen in Figure \ref{ss_rt_1s}, DU is characterized by a performance comparable to that of MVDR and MUSIC with a single snapshot, and by a good performance with multiple snapshots as we can observe in Figure \ref{ss_rt_10s}. Beside that, DU outperforms the SRP-PHAT, which is considered an effective method for reverberant environments.

\subsection{The Multiple Sources Case}

\begin{figure}[t]
\centering
\includegraphics[width=1.0\columnwidth]{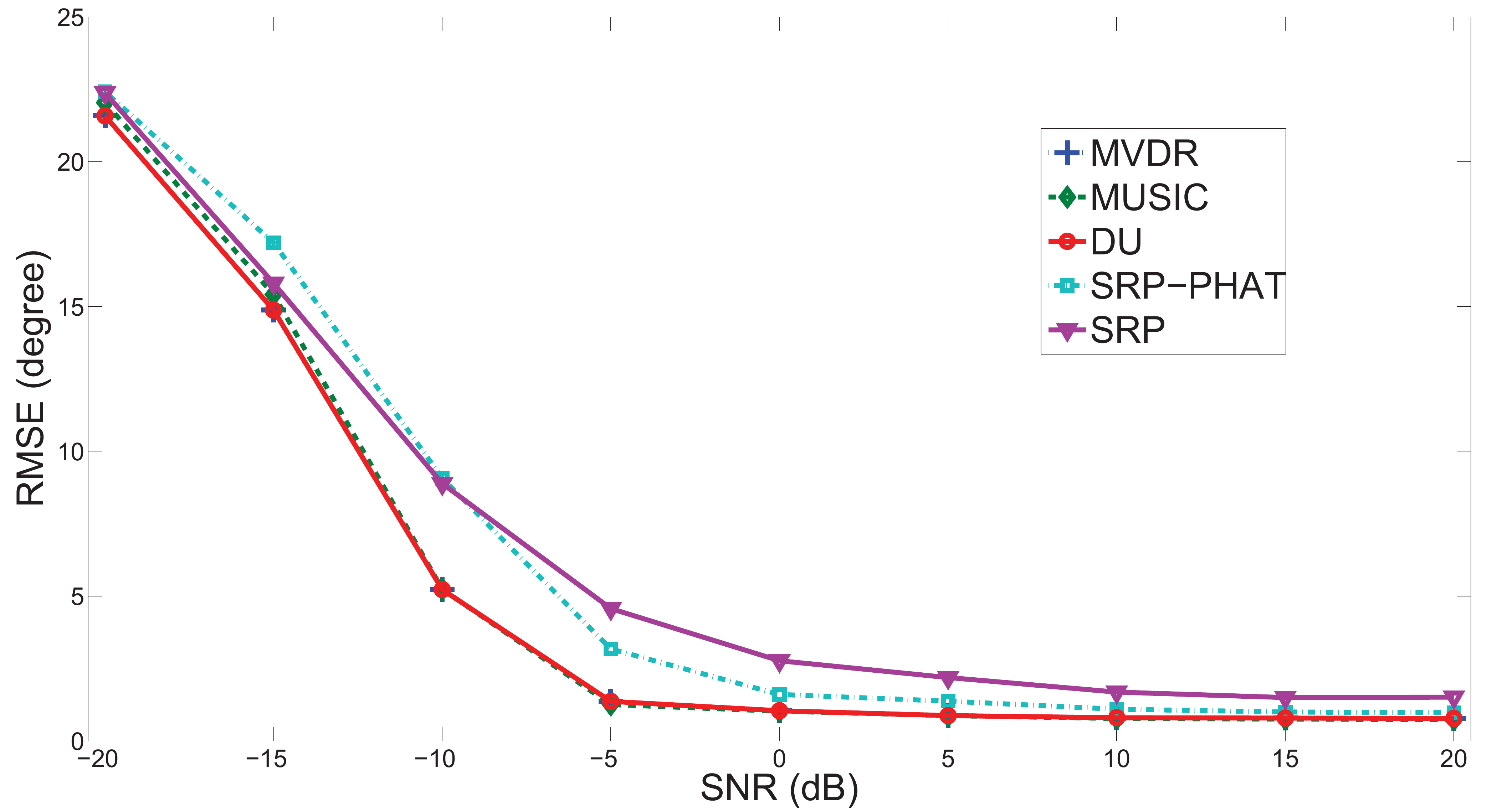}
\caption{Localization performance of two USASI noise signals in free-field at variation of SNR with a single snapshot.}
\label{ms_noise_1s}
\end{figure}

\begin{figure}[t]
\centering
\includegraphics[width=1.0\columnwidth]{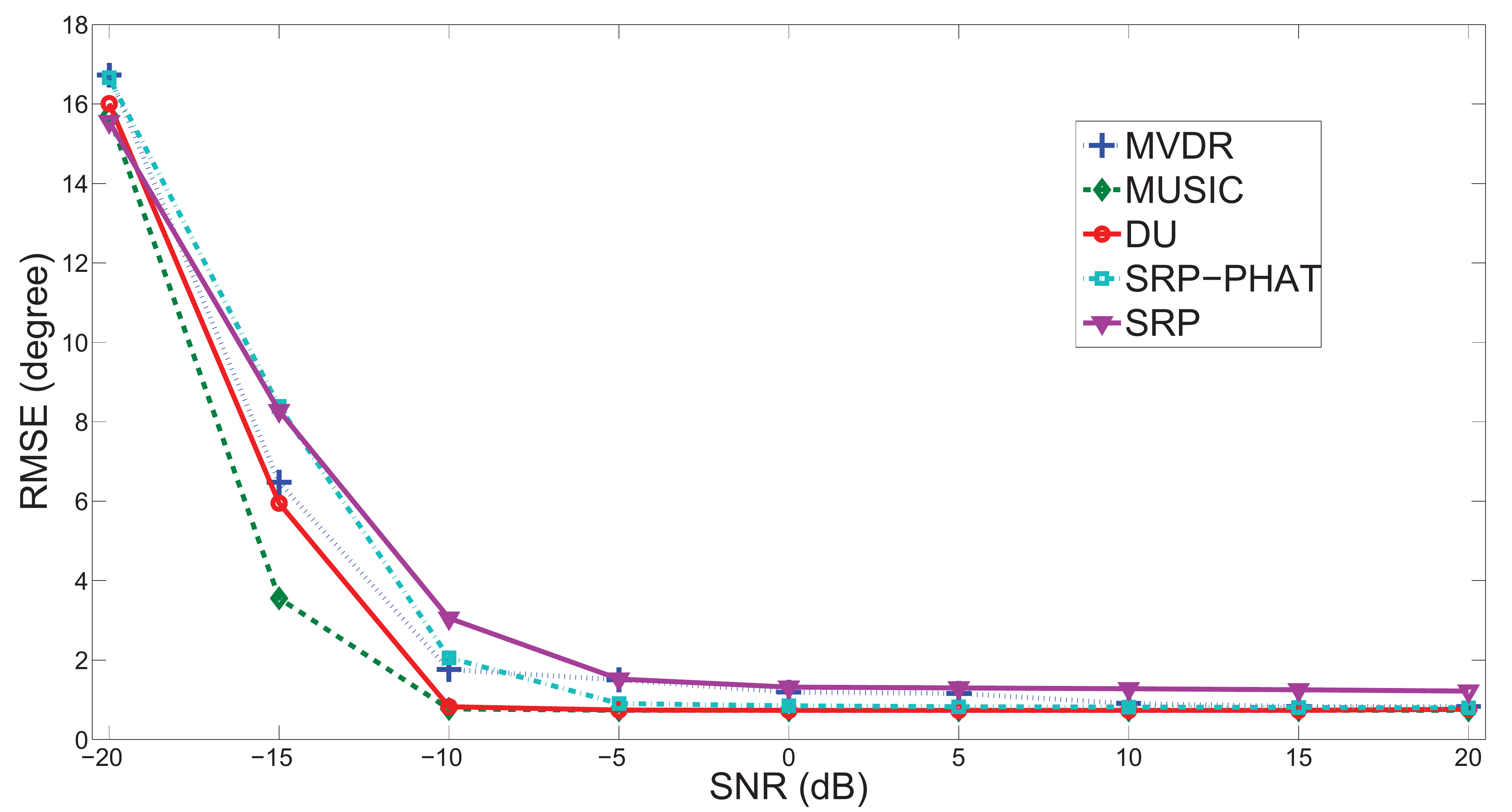}
\caption{Localization performance of two USASI noise signals in free-field at variation of SNR with a number of snapshots of 10.}
\label{ms_noise_10s}
\end{figure}

Here, we present numerical examples to verify the DOA estimation in case of two sources in the free-field case. An array of 16 sensors was used. The distance between the microphones was 0.2 m. The sources were simulated with an USASI noise signal. The sources were assumed to be stationary with a power ratio of $0.8$ between the signals and impinging the array with a $\theta_1=-11$ degree and a $\theta_2=31$ degree. The stationary nature of signals guarantees that in each frame there are always two active sources. We have assumed therefore that the noise eigenvectors have dimension $N-2$ for the MUSIC method. Note that in general the implementation of MUSIC requires that the number of sources has to be estimated \cite{WaxKailath1985}. The results at variation of SNR are depicted in Figures \ref{ms_noise_1s} and \ref{ms_noise_10s} for a single snapshot and 10 snapshots respectively. With a single snapshot the performance of DU, MUSIC, and MVDR is similar with a little degradation of MUSIC for very low SNR.   
With 10 snapshots, DU has a similar performance to that of MUSIC in the SNR range of [-10, 20] dB, whereas MUSIC outperforms DU with a SNR of -15 dB and -20 dB. In very low SNR conditions, the DU beamforming attenuation of signal eigenvectors is minor if compared to MUSIC, degrading thus the DOA estimation.

\subsection{Real Data}

The experiments were performed in a room of 4.5 m $\times$ 3.75 m $\times$ 3.05 m with a RT$_{60}$ of 0.4 s. The same array setup used for the simulated experiments with single source was considered (i.e., an uniform linear array of 8 sensors).
The distance between microphones was 0.07 m, the sampling frequency was 44.1 kHz, and the window size L was 2048 samples. A speech signal from a male speaker
was reproduced with a loudspeaker at a distance form the array of about 2 m with DOA$=[26,13,6,-13,-19,-26]$ degree. The loudspeaker consisted in a small standard cone with a diameter of about 5 cm, a frequency response of 90-20000 Hz, and a RMS power of 1 Watt. In each test position the loudspeaker was directed toward the center of the array.
The results reported in Table \ref{rwr} confirm the effectiveness of the proposed DU beamformer. We observe that the RMSE using a single snapshot is the same for MVDR, MUSIC, and DU. Besides that, in case of number of snapshot of 5 the DU outperforms all other algorithms. When the number of snapshot is 10, MUSIC provides the better DOA estimation, which is however very close to that of the proposed DU.

\begin{table}[t]
\renewcommand{\arraystretch}{1.4}
{\normalsize
\caption{The RMSE (degree) of localization performance with real data.}
\label{rwr}
\resizebox{1.0\columnwidth}{!}{
\centering
\begin{tabular}{@{}cccccc@{}}
\toprule
\textbf{Snapshot} & \textbf{MVDR} & \textbf{MUSIC} & \textbf{DU} & \textbf{SRP-PHAT} & \textbf{SRP} \\
1	& 12.448 &	12.448	& 12.448	& 12.546	& 15.518\\
5	& 7.568	& 6.894 &	6.667	& 7.204	& 8.319\\
10 &	5.419	& 4.962	& 4.980	& 5.338	& 5.255\\
\bottomrule
\end{tabular}}}
\end{table}

\section{Conclusions}
\label{sec6}

We have proposed a data-dependent DU beamformer for source localization in array signal processing. It consists of a transformation of the conventional beamformer into an high resolution method by an opportune covariance matrix conditioning operation. The DU procedure is designed to attenuate the signal subspace in the regularized covariance matrix. We have highlighted the role of the eigenvalues as weights for the attenuation of the signal eigenvectors in the calculation of the response power on one hand and, on the other, for the amplification of noise subspaces. We have introduced an eigenvalue analysis for a clear understanding of DU, MVDR and MUSIC and their relationships from the point of view of the regularized covariance matrices. The theoretical DU derivation assumes a single source and a noise-free scenario, and it is based on a minimization problem of the conventional beamforming by imposing two additional constraints on the covariance matrix. The first constraint requires that the regularized covariance matrix is negative definite, and the second one imposes that the eigenvalue corresponding to the signal subspace is zero. The latter determines how to exactly calculate the penalty weight in the DU operation. We have then analyzed the ideal DU beamforming in a more realistic noise and multisource scenario, and it has been demonstrated through simulations in different conditions that the proposed DU offers an attractive alternative to the current high resolution state-of-the-art beamformers, even though it runs at a computational cost comparable to that of the conventional beamformer. Although the MUSIC beamformer represents the best implementation for exploiting the subspace orthogonality property, the DU may have some advantages over it. MUSIC requires that the covariance matrix is accurately estimated and furthermore the number of sources has to be estimated from the eigenvalue analysis, and this operation may not be trivial in some cases. On the other hand, it has been shown that
DU is robust with respect to errors in the covariance matrix estimation, and its performance does not depend on the heuristic determination of critical parameters, as it is the case of the penalty weight for the regularization of MVDR.


\bibliographystyle{IEEEtran}
\bibliography{du}

\end{document}